\documentclass[graybox]{svmult}

\usepackage{mathptmx}       
\usepackage{helvet}         
\usepackage{courier}        

\usepackage{makeidx}         
\usepackage{graphicx}        
\usepackage{multicol}        
\usepackage[bottom]{footmisc}
\usepackage{url}

\usepackage{bm}

\begin{document}

\title*{Bayesian Inference for Continuous Time Animal Movement Based on Steps and Turns}
\author{Alison Parton, Paul G. Blackwell and Anna Skarin}
\institute{Alison Parton \at School of Mathematics and Statistics, University of Sheffield, UK, \email{aparton2@shef.ac.uk}
\and Paul G. Blackwell \at School of Mathematics and Statistics, University of Sheffield, UK, \email{p.blackwell@shef.ac.uk}
\and Anna Skarin \at Department of Animal Nutrition and Management, Swedish University of Agricultural Sciences, Sweden, \email{anna.skarin@slu.se}}

\maketitle

\abstract{Although animal locations gained via GPS, etc. are typically observed on a discrete time scale, movement models formulated in continuous time are preferable in order to avoid the struggles experienced in discrete time when faced with irregular observations or the prospect of comparing analyses on different time scales. A class of models able to emulate a range of movement ideas are defined by representing movement as a combination of stochastic processes describing both speed and bearing. \newline\indent
A method for Bayesian inference for such models is described through the use of a Markov chain Monte Carlo approach. Such inference relies on an augmentation of the animal's locations in discrete time that have been observed with error, with a more detailed movement path gained via simulation techniques. Analysis of real data on an individual reindeer {\it Rangifer tarandus} illustrates the presented methods.}

\section{Introduction}
\label{sec:introduction}
Movement ecology is a fast growing area of research concerned with addressing questions of patterns in animal movements, their underlying mechanisms, driving causes and constraints~\cite{Minerva}. Animal movement data gained via GPS, etc. are commonly given as 2-dimensional locations at a sequence of discrete---but not necessarily regular---points in time. A widespread group of models for analysing such data are based on parametrising movement by turning angles and step lengths (see e.g.~\cite{Langrock2012, McClintock2012, Morales2004, Patterson2008}). This movement is formulated in discrete time, with each subsequent location defined by a `turn' and `step' following some circular and positive distribution, respectively.

Discrete time `step and turn' models are intuitive to ecologists and statistical inference given observed data can be carried out using a range of existing methods and software. The reliance on a discrete time scale, however, poses a number of issues. The chosen scale must be ecologically relevant to the movement decisions of the animal, but is more often dictated by the sampling rate of the received data. These models are not invariant under a change of time scale, leading to no guarantee of a coherent extension to a different time scale, or how to interpret such an extension. Irregular or missing data can therefore be difficult to model, and there is often no way to compare multiple analyses defined on differing scales. 

Movement models that are formulated in continuous time avoid the discrete time difficulties; the true underlying mechanism of continuous movement is maintained, no user-defined time frame is needed and flexibility is introduced by time scale invariance. The following introduces a continuous time approach to modelling that preserves the familiar description of movement based on `steps' and `turns'.

\section{The continuous time model}
\label{sec:model}
At any time $t\geq 0$, let the animal have a bearing $\theta(t)$ and a speed $\psi(t)$ that evolve according to the stochastic differential equations
\begin{eqnarray}
\D\theta(t) = F_1 \left(t,\theta(t)\right) \D t + F_2 \left(t,\theta(t)\right) \D W_1(t), \nonumber \\
\D\psi(t)   = F_3 \left(t,\psi(t)\right)  \D t + F_4 \left(t,\psi(t)\right)   \D W_2(t),
\label{eq:contmodel}
\end{eqnarray}
where $W_i(t), \ i \in \lbrace 1,2 \rbrace$ is Brownian motion and $F_i(t,\cdot), \ i \in \lbrace 1,\ldots,4\rbrace$ are known functions.

Many discrete time `step and turn' models make the assumption that animals move with persistence, using a correlated random walk to reflect this. Such an assumption can be made within this continuous time framework by assuming $\theta(t)$ follows Brownian motion with volatility $\sigma_B^2$ by taking $F_1(t,\theta(t))=0$ and $F_2(t,\theta(t))=\sigma_B$. Note that although the direction the animal is facing is constrained to be within $[-\pi,\pi]$, $\theta(t)$ itself is not constrained in this way. Although discussed no further here, a range of other movement modes could be modelled under this framework, including directional bias and attraction to centres. 

A 1-dimensional Ornstein-Uhlenbeck (OU) process is assumed for $\psi(t)$ with parameters $\lbrace\mu,\lambda,\sigma_S^2\rbrace$, reflecting the idea that the animal's speed is stochastic but correlated over time, with some long-term average speed. This is achieved by taking $F_3(t,\psi(t))=\lambda(\mu-\psi(t))$ and $F_4(t,\psi(t))=\sigma_S$. This choice is similar to \cite{Johnson2008}, in which movement is modelled by a 2-dimensional OU velocity process. In the classic examples of discrete time models, the `step' process is assumed to be independent over disjoint time intervals. Although discussed no further here, this form of movement can easily be emulated by basing $\psi(t)$ on a distance process that follows Brownian motion with drift---where the drift describes the average speed of the animal. 

The continuous time movement model can be simulated by taking an Euler approximation over the small increment $\delta t$. Given $\theta(t)$ and $\psi(t)$ at time $t\geq0$,
\begin{eqnarray}
\theta\left(t+\delta t\right)|\theta(t)\sim \textrm{N}\left( \theta(t),\sigma_B^2\delta t\right), \nonumber \\
\psi\left(t+\delta t\right) | \psi(t) \sim \textrm{N}\left(\mu+\E^{-\lambda\delta t}\left(\psi(t)-\mu\right),\frac{\sigma_S^2}{2\lambda}\left(1-\E^{-2\lambda\delta t}\right)\right).
\label{eq:model}
\end{eqnarray}
The familiar notion of a `turn' is then given by $\theta(t+\delta t)-\theta(t)$ and a `step' by $\nu(t)=\psi(t)\delta t$.

\section{Inference for the continuous time model}
\label{sec:inference}
An animal's location $\left(\vec{X},\vec{Y}\right)$ at a series of discrete times $\vec{t}$ has been observed with error. Throughout the following, observation errors are assumed to be independent and identically distributed in both space and time, following a circular bivariate Normal distribution with variance $\sigma_E^2$. The aim of the following is to describe a method for statistical inference on the movement and error parameters $\bm{\varPhi}=\left\lbrace\sigma_B^2,\mu,\lambda,\sigma_S^2,\sigma_E^2\right\rbrace$, given $\left(\vec{X},\vec{Y}\right)$. 

\begin{figure}
\includegraphics[width=\linewidth]{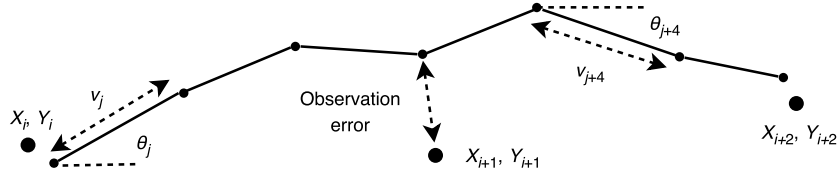}
\caption{Representation of the refined movement path $(\bm{\theta},\bm{\nu})$ and observed locations $(\vec{X},\vec{Y})$.}
\label{fig:refined_path}       
\end{figure}

It is not possible to evaluate the likelihood of $\left(\vec{X},\vec{Y}\right)$, given $\bm{\varPhi}$. The approach for inference described is to therefore augment $(\vec{X},\vec{Y})$ with a `refined path' defined by $(\bm{\theta},\bm{\nu})$. This refined path is given as a set of bearings, $\bm{\theta}$, and steps, $\bm{\nu}$, on some $\delta t$ time scale---assuming throughout that $\delta t$ is small enough that such a refined path can be taken as an approximation to the continuous time model of Eqn.~\ref{eq:contmodel}. A representation of the relationship between $\left(\vec{X},\vec{Y}\right)$ and $(\bm{\theta},\bm{\nu})$ is given in Fig.~\ref{fig:refined_path}. The joint likelihood of $\left(\vec{X},\vec{Y}\right)$ and $(\bm{\theta},\bm{\nu})$ can be evaluated, given by
\begin{equation} 
\mathcal{L} \left( \vec{X},\vec{Y},\bm{\theta},\bm{\nu} \ | \ \bm{\varPhi} \right) 
= \mathcal{L} \left( \bm{\theta},\bm{\nu} \ | \ \bm{\varPhi} \right) 
\mathcal{L} \left( \vec{X},\vec{Y} \ | \ \bm{\theta},\bm{\nu},\bm{\varPhi} \right).
\label{eq:joint_lik}
\end{equation}
The first term on the r.h.s. of Eqn.~\ref{eq:joint_lik} is the likelihood of the refined path, given by
\begin{equation} 
\mathcal{L} \left( \bm{\theta},\bm{\nu} \ | \ \bm{\varPhi} \right)
= \pi_\theta \left( \theta_1 \ | \ \bm{\varPhi} \right)
\pi_\nu \left (\nu_1 \ | \ \bm{\varPhi} \right) 
\prod_{i=2} \pi_\theta \left( \theta_i \ | \ \theta_{i-1},\bm{\varPhi} \right) 
\pi_\nu \left( \nu_i \ | \ \nu_{i-1},\bm{\varPhi} \right),
\label{eq:path_lik}
\end{equation}	
where 
\begin{eqnarray} 
\theta_1 \ | \ \bm{\varPhi} \sim \textrm{U} (-\pi,\pi), \\
\nu_1 \ | \ \bm{\varPhi} \sim \textrm{N}\left(\delta t \mu, \frac{\delta t^2 \sigma_S^2}{\lambda} \right),
\end{eqnarray}	
and $\pi_\theta( \theta_i \ | \ \theta_{i-1},\bm{\varPhi}),\pi_\nu(\nu_i \ | \ \nu_{i-1},\bm{\varPhi} )$ are given by Eqn.~\ref{eq:model} for $i \geq 2$. The second term on the r.h.s. of Eqn.~\ref{eq:joint_lik} is the likelihood of the observation error when $(\bm{\theta},\bm{\nu})$ is the `true' path.

The refined path is unknown and is simulated using a Metropolis-within-Gibbs sampler to carry out full inference on the movement parameters. This sampler alternately updates $\bm{\varPhi}$ and $(\bm{\theta},\bm{\nu})$, both also conditioned on $(\vec{X},\vec{Y})$. The respective full conditional distributions cannot be directly sampled from, and so the Metropolis-Hastings (MH) algorithm is used within each of the two steps.

\subsection{Approach for sampling the movement parameters}
\label{subsec:param}
The full conditional distribution of the movement parameters $\bm{\varPhi}$, given the refined path $(\bm{\theta},\bm{\nu})$ and the observed positions $(\vec{X},\vec{Y})$ is given by 
\begin{eqnarray}
\mathcal{L} \left( \bm{\varPhi} \ | \ \bm{\theta},\bm{\nu},\vec{X},\vec{Y}\right)
\propto \pi_\varPhi \left( \bm{\varPhi} \right)
\mathcal{L} \left( \vec{X},\vec{Y},\bm{\theta},\bm{\nu} \ | \ \bm{\varPhi}\right), 
\label{eq:param_cond_lik}
\end{eqnarray}
where $\mathcal{L}(\vec{X},\vec{Y},\bm{\theta},\bm{\nu} \ | \ \bm{\varPhi})$ is given in Eqn.~\ref{eq:joint_lik} and $\pi_\varPhi(\bm{\varPhi})$ is the prior distribution of the movement parameters. The movement parameters are proposed within the MH algorithm simultaneously using independent Normal random walks (truncated below at zero and centred on the current realisation). Acceptance is then based on the standard MH ratio using Eqn.~\ref{eq:param_cond_lik}.

\subsection{Approach for sampling the refined path}
\label{subsec:path}
The augmentation of refined movement paths is complicated by observed locations. Forward simulation based only on movement parameters will be unlikely to agree well with observations, proving infeasible within a MH step. The following describes a simulation method that, in part, takes the observations into account.

The full conditional distribution of the refined path $(\bm{\theta},\bm{\nu})$, given the movement parameters $\bm{\varPhi}$ and the observed positions $(\vec{X},\vec{Y})$ can be expressed as 
\begin{eqnarray}
\mathcal{L} \left(\bm{\theta},\bm{\nu}  \ | \ \bm{\varPhi},\vec{X},\vec{Y}\right)
= \mathcal{L} \left(\bm{\theta} \ | \ \bm{\varPhi},\vec{X},\vec{Y}\right)
 \mathcal{L} \left(\bm{\nu} \ | \ \bm{\theta}, \bm{\varPhi},\vec{X},\vec{Y}\right)
 \nonumber \\
\propto \mathcal{L} \left(\bm{\theta} \ | \ \bm{\varPhi}\right)
 \mathcal{L} \left(\vec{X},\vec{Y} \ | \ \bm{\theta},\bm{\varPhi}\right) 
 \mathcal{L} \left(\bm{\nu} \ | \ \bm{\theta}, \bm{\varPhi},\vec{X},\vec{Y}\right),
\label{eq:path_cond_lik}
\end{eqnarray}
where $\mathcal{L} \left(\bm{\theta} \ | \ \bm{\varPhi}\right)$ is given by the product of $\pi_\theta(\cdot)$ in Eqn.~\ref{eq:model}. Each observed location $(X_i,Y_i)$ can be expressed as 
\begin{eqnarray}
X_i = X_0 + \sum_j \nu_j \cos(\theta_j) + \varepsilon_{i,x}, \nonumber \\
Y_i = Y_0 + \sum_j \nu_j \sin(\theta_j) + \varepsilon_{i,y}, 
\label{eq:locs}
\end{eqnarray}
which, given $\bm{\theta}$, are linear combinations of the Normally distributed $\bm{\nu},\bm{\varepsilon}$, and so $\left(\vec{X},\vec{Y} \ | \ \bm{\theta},\bm{\varPhi}\right)$ is Normally distributed with known mean and variance. The final term in Eqn.~\ref{eq:path_cond_lik} is obtained by taking the Normally distributed $\left(\bm{\nu} \ | \ \bm{\theta}, \bm{\varPhi} \right)$, with likelihood given by $\pi_\nu(\cdot)$, and conditioning this on $\left(\vec{X},\vec{Y} \ | \ \bm{\theta},\bm{\varPhi}\right)$. The mean and variance of $\left(\bm{\nu} \ | \ \bm{\theta}, \bm{\varPhi},\vec{X},\vec{Y}\right)$ are therefore given by the standard results for multivariate conditioned Normal distributions.

Within the MH algorithm, a refined path proposal is made by first proposing bearings with density proportional to $\mathcal{L} \left(\bm{\theta} \ | \ \bm{\varPhi}\right)$. Conditional on both these bearings and observed locations, steps are proposed with density proportional to $\mathcal{L} \left(\bm{\nu} \ | \ \bm{\theta}, \bm{\varPhi},\vec{X},\vec{Y}\right)$. Acceptance of a simulated refined path is then based only on $\mathcal{L} \left(\vec{X},\vec{Y} \ | \ \bm{\theta},\bm{\varPhi}\right)$, by Eqn.~\ref{eq:path_cond_lik} and the standard MH acceptance ratio.

Proposing an entire refined path in this way is likely to yield a very low acceptance rate due to the high dimensionality. In reality, only sections of the refined path are updated at a time. This is carried out as above, but with additional conditioning upon the fixed bearings, steps and locations at the endpoints of the chosen section---i.e. $\pi_\theta(\cdot)$ is given by a Brownian bridge and $\pi_\nu(\cdot)$ is given by an OU bridge. The additional condition that the chosen section will need to meet its fixed end locations leads to the step proposal distribution being singular, and so realisations are proposed using singular value decomposition.

\section{Reindeer movement example}
\label{sec:example}
\begin{figure}
\includegraphics[width=\linewidth]{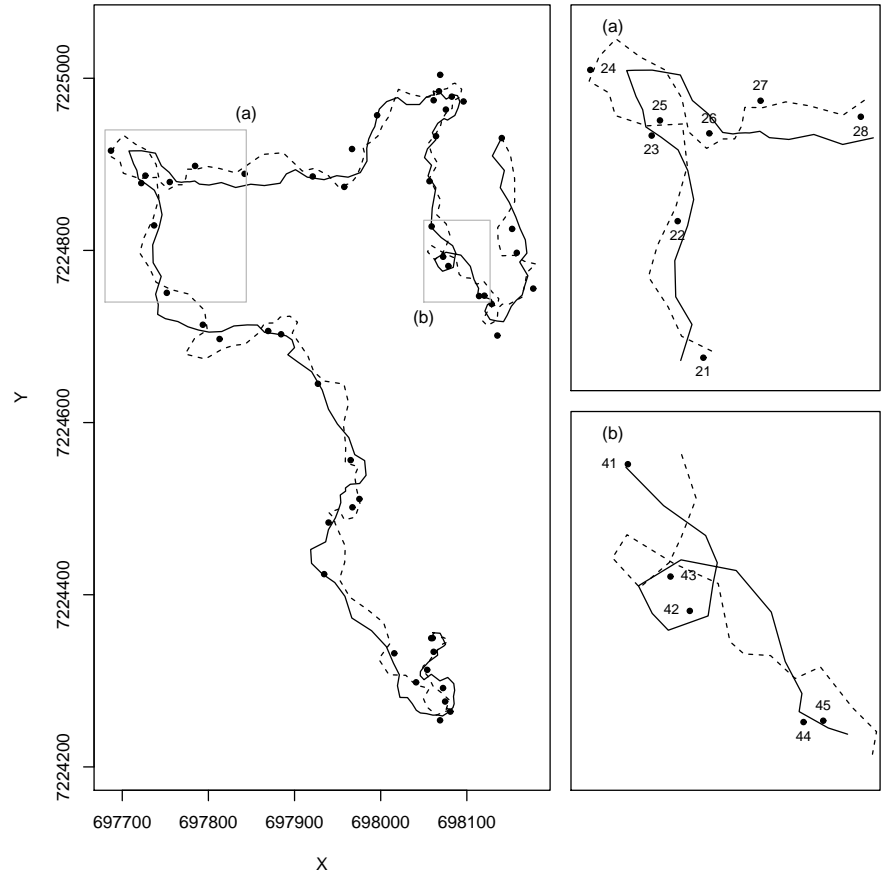}
\caption{Observations of reindeer `b53.10' (\textit{points}) and examples of two sampled paths (\textit{block and dashed lines}). Sub-figures (\textbf{a}) and (\textbf{b}) show zoomed in sections of the path, indicated by the grey boxes, with numbering showing the temporal ordering of observations.}
\label{fig:example_paths}       
\end{figure}

The method described above for statistical inference is demonstrated using observations of {\it Rangifer tarandus} movement. A subset of 50~observations of the reindeer `b53.10' walking in the Mal{\aa} reindeer herding community in northern Sweden was used, taken at mostly 2~minute intervals and shown as the points in Fig.~\ref{fig:example_paths}, with a refined path defined on a time scale of 0.5~minutes. The inference method described above was carried out with flat priors for all parameters apart from a dependence between the speed parameters to reduce the possibility of negative speeds. The refined path was sampled in short sections of between 5--12 points chosen randomly from the entire refined path, with 50 updates to the path for every parameter update in the Gibbs sampler.

A burn-in time of $10^5$~iterations was discarded and the sampler run for a further $10^5$~iterations, thinned by a factor of $10^2$. Posterior $90\%$ credible intervals for the remaining $10^3$~samples of $\bm{\varPhi}$ are given as $\sigma_B^2:(0.670,1.53), \ \mu:(24.2, 29.3), \ \lambda:(0.465,0.668), \ \sigma_S^2:(116.4,135.4), \ \sigma_E^2:(80.4,100.9)$. The posterior credible interval for $\sigma_E^2$ agrees well with current levels of observation error, expected to be up to 20~m. 

Examples of two sampled paths from throughout the run are shown in Fig.~\ref{fig:example_paths}. The marked difference in the reconstruction between some pairs of observations exhibited by the example sampled paths suggests that the linear interpolation employed by discrete time methods could be ignoring important characteristics of movement. Furthermore, in sub-plots (a) and (b) there are a number of `sharp' turns between observations 23--25 and 42--43 that have been `smoothed out' in the example path reconstructions. In a discrete time analysis this would amount to multiple turns of approximately $\pi$~radians, leading to large estimates of the turning volatility.

\section{Conclusion}
We have introduced a framework for modelling animal movement in continuous time based on the popular movement metrics of step lengths and turning angles. A method for statistical inference via the augmentation of a refined path that is assumed to approximate the continuous time path has been described and demonstrated on a subset of reindeer location observations. 

Parameter estimates for the proposed movement model give insight into the characteristics of an animal's movement in a form that is immediately interpretable, such as the mean speed at which the animal travels. These judgements are useful in addressing ecological questions relating to space/resource use, such as the size of an animal's `home range'. The augmentation method employed further supports accessible inference by supplying reconstructions of the movement path at a finer time scale than the observations. Therefore, the space use of the animal at the local scale can immediately be estimated and this enables its combination with environmental covariates, such as land cover data, whose resolution is fast increasing. The interpretation of the estimated parameters is also furthered by the ability to visualise the actual movement paths they describe.

The method here assumes a simplistic model for observation error, being Normally distributed and independent through time. A common feature of telemetry data is autocorrelation in observation error, and so in further applications more realistic models for observation error will be sought that account for this feature. 

In all of the work presented here, movement has been assumed to follow a single behavioural mode, which is unrealistic in practice for animal tracks covering an extended period of time. Behavioural switching for this model in continuous time is currently being implemented based on the works of \cite{Blackwell2015,Harris2013}, allowing switching between a finite number of `behavioural states' that represent quantitative or qualitative differences in movement.

\end{document}